\documentclass[a4paper]{article}
\usepackage{graphicx}
\usepackage{amsmath}

\begin{document}

\title{The condition for entanglement enhanced information transmission of Pauli
memory channel }
\author{Xiao-yu Chen \\
Lab. of quantum information, China Institute of Metrology, Hangzhou, 310034,
China}
\date{}
\maketitle

\begin{abstract}
The sufficient condition of entanglement enhanced classical capacity is
given for Pauli memory channel with arbitrary channel parameters. In some
special case the condition is also necessary but fail to be necessary in
general. The theory of majorization and perturbation are used in the proving.
\end{abstract}

\section{Introduction}

Calculating the capacities of quantum channels is an important task of
quantum information theory. Besides the uncorrelated channel of qubit
system, aiming at finding the entanglement enhanced channel capacity, the
memory channel has been widely discussed \cite{Macchiavello0} \cite{Yeo}
\cite{Macchiavello} \cite{Bowen1} \cite{Ball} \cite{Banaszek} \cite{Bowen2}
\cite{Giovannetti} \cite{Kretschmann} \cite{Arshed}. Among which the is
Markov channel. The simplest situation is Pauli memory channel, which
transform the state in the fashion of \cite{Macchiavello}
\begin{equation}
\rho \rightarrow \sum_{i,j=0}^3p_{ij}\sigma _i\otimes \sigma _j\rho \sigma
_i\otimes \sigma _j  \label{wave}
\end{equation}
where $p_{ij}=(1-\mu )q_iq_j+\mu q_i\delta _{ij},$ $\sigma _0=I_2,\sigma _i$
$(i=1,2,3)$ are Pauli matrices. Some of the special channels such as the
channel with $q_1=q_2=q_3=(1-x)/3,q_0=x$ and the channel with $q_0=q_1=x,$ $%
q_2=q_3=\frac 12-x$ have been studied throughoutly, and the conditions for
entangled states maximizing the channel capacity have been obtained. While
for the general single qubit channel parameters $q_i$, the problems that if
the channel capacity can be maximized by entangled state and at what
condition it can be maximized are not known.

For a general quantum channel $\mathcal{E}$, the output state is $\mathcal{E}%
(\rho )$ where $\rho $ is the input state. Holevo quantity is defined as $%
\chi (\mathcal{E})=S(\mathcal{E}(\sum_ip_i\rho _i))-\sum_ip_iS(\mathcal{E}%
(\rho _i)),$ where $S(\varrho )$ $=-$Tr$(\varrho \log _2\varrho )$ is the
von Neumann entropy of the density operator $\varrho ,$ the input ensemble is%
$\{p_i,\rho _i\}$ where $\rho _i$ are the input state on which classical
information is encoded, and are transmitted with prior probabilities $p_i.$
The channel capacity is the maximization of Holevo quantity over all input
ensembles $\{p_i,\rho _i\}$\cite{Holevo} \cite{Schumacher}$.$
\begin{equation}
C=\max_{\{p_i,\rho _i\}}\chi (\mathcal{E}).
\end{equation}
In the memory Pauli channel (\ref{wave})(we hereafter denote it as $\mathcal{%
E}$), the $\rho _i$ describe state of two qubits, thus the capacity is the
two-qubit capacity of the channel. Due to the symmetry of the channel, the
capacity was simplified to \cite{Macchiavello}
\begin{equation}
C=2-\min_\rho S(\mathcal{E}(\rho )).
\end{equation}
Also it is proven by the concavity of the von Neumann entropy that the input
state $\rho $ achieving the capacity should be pure state.

\section{Bell basis representation and strictly solvable channel}

The extremal input state that achieving the capacity can be expressed in
Bell basis:
\begin{equation}
\left| \psi \right\rangle =a_0\Phi ^{+}+a_1\Psi ^{+}+a_2\Psi ^{-}+a_3\Phi
^{-},
\end{equation}
then in Bell basis representation, one has $\mathcal{E}(\left| \psi
\right\rangle \left\langle \psi \right| )=B.$ The $4\times 4$ Hermite matrix
$B$ has its elements
\begin{eqnarray}
B_{ii} &=&(A_0,A_1,A_2,A_3)U_i(\left| a_0\right| ^2,\left| a_1\right|
^2,\left| a_2\right| ^2,\left| a_3\right| ^2)^T, \\
B_{ij} &=&a_ia_j^{*}A_{ij}+a_i^{*}a_jA_{ij}^{\prime },\text{ \ }i\neq j
\end{eqnarray}
where $U_0=\sigma _0\otimes \sigma _0,$ $U_1=\sigma _0\otimes \sigma
_1,U_2=\sigma _1\otimes \sigma _0,U_3=\sigma _1\otimes \sigma _1.$ $%
A_{ij}=A_{ji}=A_{\overline{ij}},$ ($\overline{ij}$ is the subscribes other
than $ij,$ i.e.$A_{\overline{01}}=A_{23}$ ) $A_{ij}^{\prime }=A_{ji}^{\prime
}=A_{\overline{ij}}^{\prime },$ Denote $A_4=A_{01},A_6=A_{02},A_8=A_{03};$ $%
A_5=A_{01}^{\prime },A_7=A_{02}^{\prime },A_9=A_{03}^{\prime }.$ With $%
A_0=\sum_ip_{ii},$ $A_{1,5}=2(p_{01}\pm p_{23}),$ $A_{2,7}=2(p_{02}\pm
p_{13}),$ $A_{3,9}=2(p_{03}\pm p_{12}),$ $A_4=p_{00}+p_{11}-p_{22}-p_{33},$ $%
A_6=p_{00}-p_{11}+p_{22}-p_{33},$ $A_8=p_{00}-p_{11}-p_{22}+p_{33}.$

The completely solvable channel is the channel that has been considered by
Macchiavello \cite{Macchiavello}: the channel with $q_0=q_1=x,$ $%
q_2=q_3=\frac 12-x.$ With our Bell basis representation, we will give rise
to a different way with that of Macchiavello of solving the problem. For
this channel, we have $A_5=A_6=A_8=A_9=0,$ and $A_2=A_3.$ Thus the $4\times
4 $ Hermite matrix $B$ reduces to two $2\times 2$ matrices, $B=B_1\oplus
B_2, $ with
\begin{equation}
B_1=\left[
\begin{array}{ll}
(A_0-A_2)\left| a_0\right| ^2+(A_1-A_2)\left| a_1\right| ^2+A_2, &
A_4a_0a_1^{*}+A_5a_0^{*}a_1 \\
A_4a_0^{*}a_1+A_5a_0a_1^{*}, & (A_1-A_2)\left| a_0\right| ^2+(A_0-A_2)\left|
a_1\right| ^2+A_2
\end{array}
\right] ,
\end{equation}
$B_2$ can be obtained by substituting $a_0$ with $a_2$ and $a_1$ with $a_3.$
Let $a_0=k\cos \theta ,$ $a_1=k\sin \theta $ $e^{i\varphi },(k\leq 1)$
without lose of generality we set $k\geq 1/2.$ The two eigenvalues of $B_1$
will be
\begin{eqnarray}
\lambda _{0,1} &=&\frac 12[2A_2+k(A_0+A_1-2A_2) \\
&&\pm k\sqrt{(A_0-A_1)^2\cos ^22\theta +\sin ^22\theta
(A_4^2+A_5^2+2A_4A_5\cos 2\varphi )}],  \nonumber
\end{eqnarray}
A similar expression for the two eigenvalues $\lambda _{2,3}$ of $B_2$ is
immediately obtained. To derive the extremal input state, we need the
following lemma,

Lemma: \textit{for any quantum state, if the difference of two of its
eigenvalues increases while keeping all the other eigenvalues invariant, the
entropy of the state will decrease. }

Proof: suppose the two eigenvalues are $\lambda _i,\lambda _j$ with $\lambda
_i>\lambda _j,$ the two eigenvalues contribute to the entropy as $%
S_{ij}=-\lambda _i\log _2\lambda _i-\lambda _j\log _2\lambda _j,$ denote $%
x=\lambda _i/(\lambda _i+\lambda _j),$ then $z>1/2.$ One has $%
S_{ij}=(\lambda _i+\lambda _j)[H_2(z)-\log _2(\lambda _i+\lambda _j)].$ The
binary entropy function $H_2(z)=-z\log _2z-(1-z)\log _2(1-z)$ is a monotonic
decreasing function of $z$ for $z>1/2.$ Thus $S_{ij}$ decreases as the
bigger eigenvalue $\lambda _i$ increasing. The entropy of the state
decreases when only two the eigenvalues become more apart while the other
eigenvalues are kept invariant.

The eigenvalues of $B_1$ can be made apart as possible while keeping that of
$B_2$. Then we deal with $B_2$ in the same manner. When $A_0-A_1>\left|
A_4\right| +\left| A_5\right| ,$ which can be simplified as
\begin{equation}
\mu >\left| 4x-1\right| ,  \label{wave1}
\end{equation}
we have
\begin{eqnarray}
\lambda _{0,1} &=&kA_{0,1}+(1-k)A_2, \\
\lambda _{2,3} &=&(1-k)A_{0,1}+kA_2.
\end{eqnarray}
Now $A_2=4(1-\mu )(\frac 12-x)x,A_1=2(1-\mu )[(\frac 12-x)^2+x^2],A_0=\mu
+A_1.$ Thus $A_0>A_1\geq A_2.$ For $k\geq \frac 12,$ we have $\lambda
_0>\lambda _1\geq \lambda _3,\lambda _0\geq \lambda _2>\lambda _3.$ If $%
\lambda _1\geq \lambda _2,$ the descending order of the eigenvalues will be $%
\lambda =(\lambda _0,\lambda _1,\lambda _2,\lambda _3)^{\downarrow
}=(\lambda _0,\lambda _1,\lambda _2,\lambda _3),$ otherwise $\lambda
=(\lambda _0,\lambda _2,\lambda _1,\lambda _3).$ The Bell state input
corresponds to the case of $k=1$ whose descending order of the eigenvalues
are $\lambda ^e=(A_0,A_1,A_2,A_2).$ The majorization of the eigenvalues
implies the minorization of the entropy, that is if $\lambda \prec \lambda
^e,$ then $S(\lambda ^e)<S(\lambda )$ \cite{Nielsen}$.$ When $\lambda _1\geq
\lambda _2,$ we have $\lambda _0-A_0=(1-k)(A_2-A_0)<0,$ $\lambda _0+\lambda
_1-A_0-A_1=(1-k)(2A_2-A_0-A_1)<0,$ $\lambda _0+\lambda _1+\lambda
_2-A_0-A_1-A_2=-(1-k)A_1<0;$ When $\lambda _1<\lambda _2,$ we have $\lambda
_0+\lambda _2-A_0-A_1=A_2-A_1\leq 0.$ Thus we proved $\lambda \prec \lambda
^e$, the extremal input state will be one of the four Bell states. The
entropy of the output is $S(\lambda ^e).$ The entanglement enhanced
condition (\ref{wave1}) is valid for all interval $x\in [0,\frac 12]$

While for the situation of $A_0-A_1<\left| A_4\right| +\left| A_5\right| $ ,
the eigenvalues of $B$ matrix are
\begin{eqnarray}
\lambda _{0,1} &=&kA_{0,1}^{\prime }+(1-k)A_2, \\
\lambda _{2,3} &=&(1-k)A_{0,1}^{\prime }+kA_2.
\end{eqnarray}
Where $A_{0,1}^{\prime }=\frac 12[A_0+A_1\pm (\left| A_4\right| +\left|
A_5\right| )].$

For $k\geq \frac 12,$ we also have $\lambda _0>\lambda _1\geq \lambda
_3,\lambda _0\geq \lambda _2>\lambda _3.$ Suppose the extremal input states
are again the states with $k=1,$ the eigenvalues in descending order are $%
\lambda ^e=(A_0^{\prime },A_1^{\prime },A_2,A_2)^{\downarrow }=.$ When $%
A_1^{\prime }>A_2,$ we have $\lambda _0-$ $A_0^{\prime
}=(1-k)(A_2-A_0^{\prime })<0,$ $\lambda _0+\lambda _1-$ $A_0^{\prime
}-A_1^{\prime }=$ $(1-k)(2A_2-A_0-A_1)<0,$ $\lambda _0+\lambda _1+\lambda
_2- $ $A_0^{\prime }-A_1^{\prime }-A_2=\lambda _1-A_1^{\prime
}=(1-k)(A_2-A_1^{\prime })<0$ for the case of $\lambda _1\geq \lambda _2$;
and $\lambda _0+\lambda _2-$ $A_0^{\prime }-A_1^{\prime }=A_2-A_1^{\prime
}<0 $ for the case of $\lambda _1<\lambda _2.$

When $A_1^{\prime }<A_2,$ we have $\lambda _0+\lambda _1-$ $A_0^{\prime
}-A_2=\lambda _0+\lambda _1-$ $A_0^{\prime }-A_1^{\prime }-(A_2-A_1^{\prime
})<0,$ $\lambda _0+\lambda _1+\lambda _2-$ $A_0^{\prime }-2A_2=k(A_1^{\prime
}-A_2)<0$ for the case of $\lambda _1\geq \lambda _2$; and $\lambda
_0+\lambda _2-$ $A_0^{\prime }-A_2=0$ for the case of $\lambda _1<\lambda
_2. $

The entanglement enhanced capacity condition (\ref{wave1}) is proved.

\section{The general Pauli channel}

Usually, for general Pauli channels, the eigenvalues as well as the entropy
of output state $\mathcal{E}(\left| \psi \right\rangle \left\langle \psi
\right| )$ can not be analytically obtained. With the numerical steepest
descending method, the extremal states can be found to be either with only
one Bell state, or the superposition of two Bell states with the same
probabilities (separable state). We will prove that such input states are
extremal by perturbation theory in the next section.

To compare the maximal of entanglement input and that of separable state
input in order to obtain the capacity, we first need to regularize the
channel. The channel $\mathcal{E}$ possesses the property that
\begin{equation}
\mathcal{E}(\sigma _i\otimes \sigma _j\rho \sigma _i\otimes \sigma
_j)=\sigma _i\otimes \sigma _j\mathcal{E}(\rho )\sigma _i\otimes \sigma _j.
\end{equation}
As entropy is invariant under unitary transformation, the input states $\rho
$ and $\sigma _i\otimes \sigma _j\rho \sigma _i\otimes \sigma _j$ have the
same output entropy. If $q_0$ is not the maximal among $q_0,q_1,q_2,q_3$,
see $q_1$ is the maximal, we use $\rho ^{\prime }=\sigma _1\otimes \sigma
_1\rho \sigma _1\otimes \sigma _1$ as the input state, the action of the
channel $\mathcal{E}$ on $\rho ^{\prime }$ will be equivalent to a channel $%
\mathcal{E}^{\prime }$ applying on $\rho $ in the sense of output entropy.
The channel $\mathcal{E}^{\prime }$ is produced from $\mathcal{E}$ by
exchanges of $q_0\leftrightarrow q_1,q_2\leftrightarrow q_3.$ The two
channels are equivalent for our problem of classical capacity. If $q_2$ or $%
q_3$ is the maximal, we can use $\sigma _2\otimes \sigma _2\rho \sigma
_2\otimes \sigma _2$ or $\sigma _3\otimes \sigma _3\rho \sigma _3\otimes
\sigma _3$as the input state. Thus the channel can be regularized to channel
with $q_0$ being the maximal among $q_0,q_1,q_2,q_3.$ Without lose of
generality, we can suppose $q_1\geq q_0,q_3.$ The Pauli channel is
regularized to
\begin{equation}
q_0\geq q_1\geq q_2,q_3.
\end{equation}
It follows immediately that $A_0\geq A_1\geq A_2,A_3$ and $A_4\geq 0,$ $%
A_5\geq 0.$ The numerical steepest descending calculation exhibits that the
extremal input pure state will be with the form of $a_0=a_1=0$ or $%
a_2=a_3=0. $ Thus the matrix $B$ is reduced to a $2\times 2$ matrix and two
diagonal elements. When $a_2=a_3=0,$
\begin{equation}
B=\left[
\begin{array}{ll}
A_0\left| a_0\right| ^2+A_1\left| a_1\right| ^2, &
A_4a_0a_1^{*}+A_5a_0^{*}a_1 \\
A_4a_0^{*}a_1+A_5a_0a_1^{*}, & A_1\left| a_0\right| ^2+A_0\left| a_1\right|
^2
\end{array}
\right] \oplus \lambda _2\oplus \lambda _3.
\end{equation}
with $\lambda _2=A_2\left| a_0\right| ^2+A_3\left| a_1\right| ^2,$ $\lambda
_3=A_3\left| a_0\right| ^2+A_2\left| a_1\right| ^2.$ Let $a_0=\cos \theta ,$
$a_1=\sin \theta $ $e^{i\varphi },$ and using the above lemma to eliminate $%
\varphi ,$ the eigenvalues of $B$ are
\begin{eqnarray}
\lambda _{0,1} &=&\frac 12[A_0+A_1\pm \sqrt{(A_0-A_1)^2\cos ^22\theta +\sin
^22\theta (A_4+A_5)^2}], \\
\lambda _{2,3} &=&\frac 12[A_2+A_3\pm (A_2-A_3)\cos 2\theta ].
\end{eqnarray}

If $A_0-A_1>A_4+A_5,$ we have $\lambda _1>A_1\geq \max (A_2,A_3)\geq \lambda
_{2,}\lambda _3.$ The descending order of the eigenvalues are $\lambda
=(\lambda _0,\lambda _1,\lambda _2^{\prime },\lambda _3^{\prime }),$ where
we have denoted $\lambda _{2,3}^{\prime }=\frac 12[A_2+A_3\pm \left|
(A_2-A_3)\cos 2\theta \right| ]$ without lose of generality. Suppose the
extremal state is with $\sin ^22\theta =0,$ the descending order of the
extremal eigenvalues are $\lambda ^e=(A_0,A_1,A_2^{\prime },A_3^{\prime }),$%
with $A_2^{\prime }=\max (A_2,A_3),A_3^{\prime }=\min (A_2,A_3).$ It is not
difficult to see that $\lambda \prec \lambda ^e$.

If $A_0-A_1<A_4+A_5,$the descending order of the eigenvalues are $\lambda
=(\lambda _0,\lambda _1,\lambda _2^{\prime },\lambda _3^{\prime
})^{\downarrow }.$ When $\lambda _1\geq \lambda _2^{\prime },\lambda
=(\lambda _0,\lambda _1,\lambda _2^{\prime },\lambda _3^{\prime }).$ Suppose
the extremal state is with $\cos 2\theta =0,$the descending order of the
extremal eigenvalues are $\lambda ^e=(A_0^{\prime },A_1^{\prime },\frac
12(A_2+A_3),\frac 12(A_2+A_3))$ with $A_{0,1}^{\prime }=\frac 12[A_0+A_1\pm
(A_4+A_5)].$ We have $\lambda _0<A_0^{\prime },$ $\lambda _0+\lambda
_1=A_0^{\prime }+A_1^{\prime },$ but $\lambda _2^{\prime }\geq \frac
12(A_2+A_3).$When $\lambda _1<\lambda _2^{\prime }$, the situation may even
be worse. Thus we can not conclude that $\lambda ^e$ is extremal by
majorization. Only in the situation of $A_2=A_3,$ that is $q_0=q_1$ or $%
q_2=q_3$, we can conclude that $\lambda ^e$ is extremal by majorization. For
most of channels, our numerical result exhibits that the extremal state is
with $\cos 2\theta =0,$ thus the input state is $\left| \psi \right\rangle $
$=\frac 1{\sqrt{2}}(\left| 0\right\rangle +\left| 1\right\rangle )\otimes $ $%
\frac 1{\sqrt{2}}(\left| 0\right\rangle +\left| 1\right\rangle )$ or $\frac
1{\sqrt{2}}(\left| 0\right\rangle -\left| 1\right\rangle )\otimes \frac 1{%
\sqrt{2}}(\left| 0\right\rangle -\left| 1\right\rangle ).$ There exist the
situation that Bell state maximize the capacity even when $A_0-A_1<A_4+A_5,$
although it takes place scarcely in our numerical calculation.

Thus $A_0-A_1>A_4+A_5$ is only the sufficient condition (not necessary) of
entanglement enhanced capacity. It can be written as
\begin{equation}
\mu >\frac{2q_0q_1-q_2^2-q_3^2}{q_2+q_3+2q_0q_1-q_2^2-q_3^2}.
\end{equation}

\section{Perturbation Verification}

The numerical calculation give rise to the results that the extremal input
pure state is either Bell state or superposition of two Bell states (with
equal probabilities, a separable state). In the Bell state basis, they are $%
[1,0,0,0]$ are $\frac 1{\sqrt{2}}[1,1,0,0]$ or the similar. For a
perturbation to Bell state input $\Phi ^{+},$ suppose the perturbed input is
$[a_0,a_1,a_2,a_3]$ with $\left\| 1-a_0,a_1,a_2,a_3\right\| \rightarrow 0$,
we have The matrix $B$ now can be written as $B_0+$ $B^{\prime },$ with
\begin{equation}
B_0=diag\{A_0,A_1,A_2,A_3\},
\end{equation}
The first order perturbations to the eigenvalues of $B_0$ are $%
B_{ii}^{\prime }$ (i=0,1,2,3).
\begin{equation}
B_{ii}^{\prime }=(A_0,A_1,A_2,A_3)U_i(\left| a_0\right| ^2,\left| a_1\right|
^2,\left| a_2\right| ^2,\left| a_3\right| ^2)^T-A_i.
\end{equation}
Keep in mind that $A_0\geq A_1\geq A_2,A_3,$ we have $B_{00}^{\prime
}=(\left| a_0\right| ^2-1)A_0$ $+\left| a_1\right| ^2A_1+\left| a_2\right|
^2A_2+\left| a_3\right| ^2$ $A_3=$ $\left| a_1\right| ^2(A_1-A_0)+$ $\left|
a_2\right| ^2$ $(A_2-A_0)+\left| a_2\right| ^2$ $(A_3-A_0)\leq 0;$ $%
B_{00}^{\prime }+B_{11}^{\prime }=(\left| a_0\right| ^2-1)(A_0+A_1)$ $%
+\left| a_1\right| ^2(A_0+A_1)+\left| a_2\right| ^2(A_2+A_3)$ $+$ $\left|
a_3\right| ^2(A_2+A_3)=-(\left| a_2\right| ^2+\left| a_3\right|
^2)(A_0+A_1-A_2+A_3)\leq 0.$ Suppose $A_3<A_2$, $B_{00}^{\prime
}+B_{11}^{\prime }+B_{22}^{\prime }=-B_{33}^{\prime }=A_3-(A_3\left|
a_0\right| ^2+A_2\left| a_1\right| ^2+A_1\left| a_2\right| ^2+A_3\left|
a_3\right| ^2)\leq 0.$ If $A_3>A_2,$ we get $B_{00}^{\prime }+B_{11}^{\prime
}+B_{33}^{\prime }=-B_{22}^{\prime }\leq 0.$ In either cases, the descending
order eigenvalues of $B_0$ majorizes that of $B_0+$ $B^{\prime }.$ So that
the perturbation will increase the entropy of the output state.

For a perturbation to the product input state $\frac 1{\sqrt{2}}(\left|
0\right\rangle +\left| 1\right\rangle )\otimes $ $\frac 1{\sqrt{2}}(\left|
0\right\rangle +\left| 1\right\rangle )=\frac 1{\sqrt{2}}(\Phi ^{+}+\Psi
^{+}),$ suppose the perturbed input is $[a_0,a_1,a_2,a_3]$ with $\left\| 1/%
\sqrt{2}-a_0,1/\sqrt{2}-a_1,a_2,a_3\right\| \rightarrow 0$, the unperturbed
matrix is
\begin{equation}
B_0=\frac 12\left[
\begin{array}{ll}
A_0+A_1, & A_4+A_5 \\
A_4+A_5, & A_1+A_0
\end{array}
\right] \oplus \frac 12(A_2+A_3)\oplus \frac 12(A_2+A_3).
\end{equation}
with eigenvalues $\lambda _{0,1}=\frac 12[A_0+A_1\pm (A_4+A_5)],$ $\lambda
_{2,3}=\frac 12(A_2+A_3).$ The first order perturbations to the the first
two eigenvalues of $B_0$ are
\begin{equation}
\lambda _{0,1}^{\prime }=\frac 12[-(1-\left| a_0\right| ^2-\left| a_1\right|
^2)(A_0+A_1-A_2-A_3)\pm (a_0a_1^{*}+a_0a_1^{*}-1)(A_4+A_5)].
\end{equation}
The last two eigenvalues of $B_0$ are degenerated, thus the degenerate
perturbation should be applied. The perturbation to the eigenvalues are
\begin{equation}
\lambda _{2,3}^{\prime }=\frac 12[(1-\left| a_0\right| ^2-\left| a_1\right|
^2)(A_0+A_1-A_2-A_3)\pm \sqrt{C}],
\end{equation}
where $C=[(\left| a_0\right| ^2-\left| a_1\right| ^2)(A_2-A_3)+(\left|
a_2\right| ^2-\left| a_3\right| ^2)(A_0-A_1)]^2+$ $4[\left| a_2\right|
^2\left| a_3\right| ^2(A_4^2+A_5)$ $+$ $%
(a_2^{*2}a_3^2+a_2^2a_3^{*2})A_4A_5]. $ We can easily see that $\lambda
_0^{\prime }\leq 0.$ If $A_0+A_1-A_4-A_5>A_2+A_3,$ that is
\begin{equation}
\mu >1-\frac 1{2(q_0+q_1)},  \label{we1}
\end{equation}
the descending order of eigenvalues of $B_0$ is $(\lambda _0,\lambda
_1,\lambda _2,\lambda _3).$ We need to verify that $\lambda _0^{\prime
}+\lambda _1^{\prime }\leq 0$ and $\lambda _0^{\prime }+\lambda _1^{\prime
}+\lambda _2^{\prime }=-\lambda _3^{\prime }\leq 0.$ The former is evident.
The condition $\lambda _3^{\prime }\geq 0$ should be true for all possible
perturbations. As far as $A_2\neq A_3,$we can choose the situation of $%
\left| a_2\right| =\left| a_3\right| $ $=0$ while $\left| a_0\right| \neq
\left| a_1\right| ,$ so that $\lambda _3^{\prime }<0.$ When $A_2=A_3,$the
condition for $\lambda _3^{\prime }\geq 0$ is
\begin{equation}
4(A_0-A_2)(A_1-A_2)>(A_4+A_5)^2.  \label{we2}
\end{equation}

If $A_0+A_1-A_4-A_5<A_2+A_3,$the descending order of eigenvalues of $B_0$ is
$(\lambda _0,\lambda _2,\lambda _3,\lambda _1).$ We need to verify that $%
\lambda _0^{\prime }+\lambda _2^{\prime }\leq 0$ and $\lambda _0^{\prime
}+\lambda _2^{\prime }+\lambda _3^{\prime }=-\lambda _1^{\prime }\leq 0.$
The later is evident by the fact that $A_4+A_5>A_0+A_1-A_2-A_3$ and $%
a_0a_1^{*}+a_0a_1^{*}\leq \left| a_0\right| ^2+\left| a_1\right| ^2$. The
condition $\lambda _0^{\prime }+\lambda _2^{\prime }\leq 0$ requires
\begin{equation}
(1-a_0a_1^{*}-a_0a_1^{*})(A_4+A_5)\geq \sqrt{C}.  \label{wee}
\end{equation}
When $A_2\neq A_3,$we can choose the situation of $\left| a_2\right| =\left|
a_3\right| $ $=0$ to verify that (\ref{wee}) can not be true in general. For
the situation of $A_2=A_3,$ the condition (\ref{wee}) will reduce to $%
A_4+A_5>A_0-A_1$ which is just the requirement of separable state achieving
the capacity.

Hence, when $A_2\neq A_3,$ even in the sense of perturbation, majorization
can not be used to prove the extremal property of the product state as an
input to the channel.

\section{Conclusion}

The classical capacity of Pauli channel is investigated with the
representation of Bell states. A new proof is given for the capacity of some
strictly solvable symmetric Pauli channel ($q_0=q_1$ and $q_2=q_3),$ the
full expression of the entanglement enhanced classical condition is (\ref
{wave1}) . For the most general Pauli channel, the condition for
entanglement enhanced classical capacity is given. The condition is a
sufficient condition but is not necessary. When in the situation of $A_2=A_3,
$ that is, the two big $q_i$ are equal or the two small $q_i$ are equal, the
condition is also necessary under a group of additional inequalities which
are obtained by perturbation theory. This comprise the well studied channel
of $q_1=q_2=q_3=(1-x)/3,q_0=x.$ For this channel, it should be mentioned
that the full expression of the entanglement enhanced classical capacity
condition is $\mu >\frac{\left| 4x-1\right| /3}{1+\left| 4x-1\right| /3}.$

Funding by the National Natural Science Foundation of China ( Grant No.
10575092,10347119), Zhejiang Province Natural Science Foundation (Grant No.
RC104265) and AQSIQ of China ( Grant No. 2004QK38) are gratefully
acknowledged.


\begin{thebibliography}{99}
\bibitem{Macchiavello0}  C. Macchiavello and G. M. Palma, Phys. Rev. A
\textbf{65}, 050301(R) (2002).

\bibitem{Yeo}  Y. Yeo and A. Skeen, Phys. Rev. A \textbf{67}, 064301 (2003).

\bibitem{Macchiavello}  C. Macchiavello, G. M. Palma, S. Virmani, Phys. Rev.
A, \textbf{69}, 010303(R), (2004).

\bibitem{Bowen1}  G. Bowen and S. Mancini, Phys. Rev. A \textbf{69}, 012306
(2004).

\bibitem{Ball}  J. Ball, A. Dragan and K. Banaszek, Phys. Rev. A \textbf{69}%
, 042324 (2004).

\bibitem{Banaszek}  K. Banaszek, A. Dragan, W. Wasilewski and C. Radzewicz,
Phys. Rev. Lett. \textbf{92}, 257901 (2004).

\bibitem{Bowen2}  G. Bowen, I. Devetak and S. Mancini, Phys. Rev. A \textbf{%
71}, 034310 (2005).

\bibitem{Giovannetti}  V. Giovannetti and S. Mancini, Phys. Rev. A \textbf{71%
}, 062304 (2005).

\bibitem{Kretschmann}  D. Kretschmann and R. F. Werner, Phys. Rev. A \textbf{%
72}, 062323 (2005).

\bibitem{Arshed}  N. Arshed and A. H. Toor, Phys. Rev. A \textbf{73}, 014304
(2006).

\bibitem{Holevo}  A. S. Holevo, IEEE Trans. Inf. Theory, \textbf{44}, 269
(1998).

\bibitem{Schumacher}  B. Schumacher, M. Westmoreland, Phys. Rev. A, \textbf{%
56}, 131 (1997).

\bibitem{Nielsen}  See for example: M. A. Nielsen and I. L. Chuang, Quantum
Computation and Quantum Information (Cambridge University Press, Cambridge,
2000).
\end{thebibliography}
\end{document}